\documentclass[graybox]{svmult}

\usepackage{mathptmx}       
\usepackage{helvet}         
\usepackage{courier}        
\usepackage{type1cm}        
\usepackage{makeidx}         
\usepackage{graphicx}        
\usepackage{multicol}        
\usepackage[bottom]{footmisc}
\usepackage{natbib}
\setcitestyle{aysep={}}
\def\apj{ApJ}
\def\apjl{ApJL}
\def\apjs{ApJS}
\def\mnras{MNRAS}
\def\pasa{Publ. Astron. Soc. Australia}
\def\aap{A\&A}
\def\aj{AJ}
\def\jqsrt{J.Quant.Spectrosc.Radiat.Trans.}
\def\nat{Nature}
\def\pasj{PASJ}
\def\araa{Ann. Rev. Astron. Ap.}

\makeindex             

\begin{document}

\title*{Manifestations of the Galactic Center Magnetic Field}
\titlerunning{Galactic Center Magnetic Field}
\author{Mark R. Morris}
\institute{University of California, Los Angeles}
%
%
\maketitle

\abstract{Several independent lines of evidence reveal that a relatively strong and highly ordered magnetic field is present throughout the Galaxy's central molecular zone (CMZ).  The field within dense clouds of the central molecular zone is predominantly parallel to the Galactic plane, probably as a result of the strong tidal shear in that region. A second magnetic field system is present outside of clouds, manifested primarily by a population of vertical, synchrotron-emitting filamentary features aligned with the field.  Whether or not the strong vertical field is uniform throughout the CMZ remains undetermined, but is a key central issue for the overall energetics and the impact of the field on the Galactic center arena.  The interactions between the two field systems are considered, as they are likely to drive some of the activity within the CMZ.  As a proxy for other gas-rich galaxies in the local group and beyond, the Galactic center region reveals that magnetic fields are likely to be an important diagnostic, if not also a collimator, of the flow of winds and energetic particles out of the nucleus.}

\section{Introduction}
The Central Molecular Zone (CMZ) of the Galaxy, which represents the strongest concentration of molecular gas in the Galaxy, is immersed in a magnetic field that is manifested in two rather distinct ways.  First, the field within clouds appears to be predominantly toroidal, that is, oriented parallel to the Galactic plane, while the field external to clouds or in low-density gas is largely poloidal, or perpendicular to the Galactic plane.  The following sections describe the evidence for each of these separately, and present the implications for the interactions of the two components.  Previous reviews that emphasize other aspects of the magnetic field are given by \cite{Bicknell:2001zr}, \cite{Morris:2006fk} and \cite{Ferriere:2009fk, Ferriere:2011vn}.  What has been learned from the Galactic center is probably generalizable to other barred galaxies having a substantial gas content in their central regions.  However, it is only in our Galaxy that present techniques allow us to elucidate the essential details. 

\section{The Sheared Field in the Central Molecular Zone}
The most successful way of probing the field within molecular clouds of the CMZ has been by measuring the polarization of the thermal emission from magnetically aligned dust grains \citep{Hildebrand:2002kx}.  For a wide range of interpretations of the alignment mechanism, the field direction projected onto the plane of the sky is perpendicular to the measured polarization E-vector \citep{Lazarian:2007fk}.  Measurements have been made at far-infrared \cite[e.g.,][]{Chuss:2003fk}, submillimeter \cite[e.g.,][]{Novak:2003kx}, and near-infrared \citep{Nishiyama:2010uq} wavelengths.  On large scales near the Galactic plane, the field direction is generally found to be parallel to the Galactic plane, although in detail, the field direction tends to follow the long axis of the cloud being observed \citep{Chuss:2003fk}.  This has been attributed to tidal shear of gas in the CMZ \citep{Aitken:1991vn, Aitken:1998uq, Chuss:2003fk}.  The Galactic tidal force near the center is strong enough that clouds are not stable against tidal shear unless their density exceeds a fairly large value, 10$^4$ cm$^{-3}$[75 pc/R$_{gc}]^{1.8}$  \citep{Gusten:1989uq}, where R$_{gc}$ is the galactocentric radius.  Consequently, many of the "clouds" in the CMZ are in fact best described as tidal streams, some of which subtend a large wrapping angle around the Galactic center \citep[e.g.,][]{Tsuboi:1999fk}.  This continuous shearing, coupled with flux-freezing of the magnetic field to the partially ionized molecular medium, transforms any pre-existing magnetic field geometry within the clouds into one that is dominated by a large-scale toroidal (i.e., azimuthal) component in which the field lines follow the shear.  Clouds that, for whatever local reason, are sheared in a direction at some angle to the Galactic plane, such as M0.10+0.02, the molecular cloud underlying the arched radio filaments \citep{Morris:1992zr, Chuss:2003fk}, are consistent with this picture, inasmuch as the field tends to follow the long axis of such clouds.  These points are well illustrated by Figure 2, from \cite{Chuss:2003fk}, which shows measured magnetic field vectors superimposed on a radio map.  

\begin{figure}
    \centering
        \includegraphics[width=7cm]{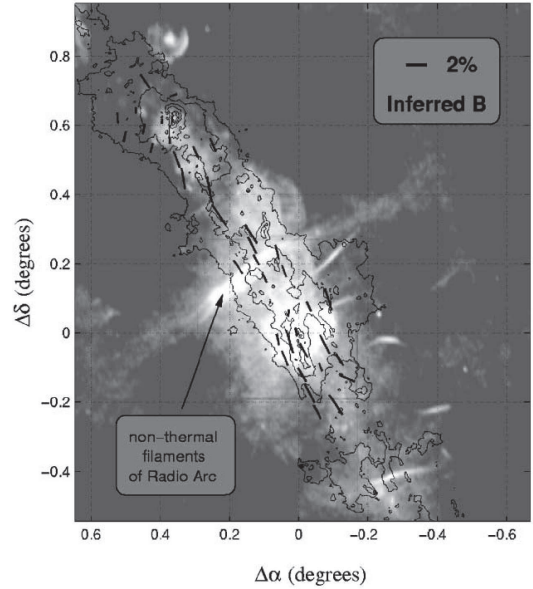}

        \vspace{-1mm}
        \caption{Magnetic field vectors inferred from polarization measurements at 450 $\mu$m by \cite{Novak:2003kx} superimposed upon a grayscale 90-cm radio continuum image from \cite{LaRosa:2000bs} and contours of 850 $\mu$m continuum emission from \cite{Pierce-Price:2000uq}} 

\end{figure}

\begin{figure}
    \centering
        \includegraphics[width=8cm]{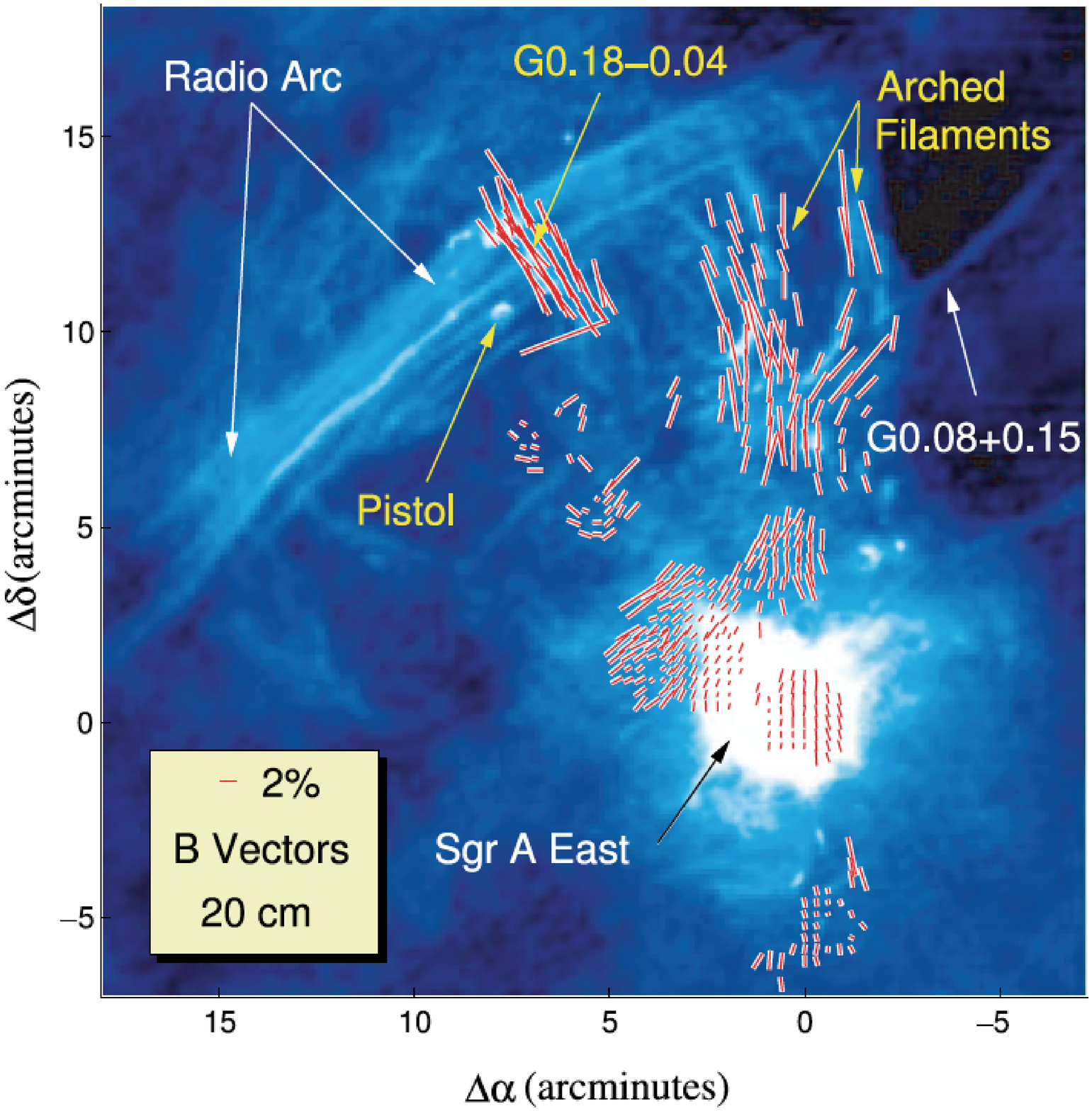}

        \vspace{-1mm}
        \caption{Magnetic field vectors derived by \cite{Chuss:2003fk} from polarization measurements, superimposed upon a 20-cm radio continuum image from \cite{Yusef-Zadeh:1984fk}.  The Galactic plane is oriented in a position angle of about 30 degrees, so that the field vectors in G0.18-0.04 are well aligned along it.  These vectors, plus the bundle of NTFs in the Radio Arc, best illustrate the orthogonality between the field inside and outside of clouds \cite[see discussion in][]{Morris:1996oq}.  Note also that the field in the arched radio filaments, which outline the surface of an underlying molecular cloud, follows the filament orientations quite closely. In addition, the indicated field directions in several low-density regions do not conform to the toroidal trend for field vectors in more massive clouds such as G0.18-0.04.} 

\end{figure}

The parsec-scale circumnuclear disk (CND) around the central black hole is an illustrative case that is consistent with the anticipated effects of shear.  Field vectors implied by the 100-$\mu$m polarization measurements of \cite{Hildebrand:1993fk} are consistent with a self-similar model by \cite{Wardle:1990ly} in which an initially vertical field threading the differentially rotating disk is pulled into an axisymmetric configuration.  The $\sim20^{\circ}$ tilt of the CND with respect to the Galactic plane is reflected in the orientation of the inferred magnetic field vectors.  The CND will be a key object for investigating the interplay between gas dynamics and the magnetic field as higher resolution polarization measurement capabilities are brought to bear on it in the future, such as with the HAWC+ instrument on SOFIA.  

The above generalizations about shearing do not necessarily apply to the densest clouds of the CMZ -- those which are presumed tidally stable. In such cases, the field geometry would reflect the peculiar dynamical history of the cloud, although shearing that took place in some of the pre-existing constituents of a dense cloud might still lead to its having a dominant toroidal field.  The field vectors in the dense M-0.13-0.08 cloud are roughly parallel to the Galactic plane \citep{Dotson:2010ve}, and might therefore provide an illustration of this last point.  The dense clouds M-0.02-0.07 and Sgr B2, on the other hand, show field vectors that are complex and generally inconsistent with an overall toroidal geometry \citep{Novak:1997qf, Dotson:2010ve}.  In the case of Sgr B2, the complexity is compounded by high optical depths in the dust emission; some of the polarization vectors in the central regions of Sgr B2 are determined by foreground absorption within the source \citep{Novak:1997qf}.  

\cite{Chuss:2003fk} find a striking relationship between dust column density, as inferred from their measured 350 µm flux, and the mean angle between the measured polarization E-vector and that which would pertain to a purely poloidal magnetic field.  They interpret this in terms of a pervasive poloidal field throughout the central few hundred parsecs of the Galactic center that becomes progressively more toroidal as the gas density increases, with the ratio of the gravitational energy density to the magnetic energy density therefore rising with density until the gravitational shear produces a toroidal magnetic field geometry in the relatively dense gas.  Using their submillimeter fluxes to obtain column densities, \cite{Chuss:2003fk} derive a strength for the poloidal magnetic field of a few milligauss\footnote{If applicable, this and the considerations in the previous paragraph imply that there is both a lower and an upper limit to the density of clouds within which the field is likely to be toroidal}.  A likely related trend was reported by \cite{Nishiyama:2010uq}.  They find from their near-infrared observations that the mean magnetic field orientation evolves smoothly from a toroidal field near the Galactic plane to a poloidal field at Galactic latitudes above 0.4 degrees.   However, in contrast to Chuss et al., and in spite of their lower spatial resolution, Nishiyama et al.\ consider that the magnetic field near the Galactic plane remains toroidal even in the intercloud medium.  Implications of the dual field geometry  are discussed further below.  
 
\section{Nonthermal Radio Filaments -- the Poloidal Field}
Over the past 30 years, the VLA has revealed a population of filamentary radio structures rising from within the CMZ.  Their emission is polarized, indicative of synchrotron emission, and when the polarized E-vectors are corrected for foreground Faraday rotation, the implied magnetic field direction is found to be aligned with the nonthermal filaments (NTFs).  Numerous ideas have been advanced to account for the NTFs \citep[references in][]{Morris:1996fk, Chevalier:1992dq, Rosner:1996bh, Shore:1999vn, Bicknell:2001zr, Yusef-Zadeh:2003uq, Boldyrev:2006fk}, but there is presently no consensus on the nature of the mechanism that has produced them.  One important clue is that the predominant orientation of the NTFs -- particularly the brighter and longer ones -- is roughly perpendicular to the Galactic plane \cite{Morris:1985kl, Anantharamaiah:1991ve, Nord:2004fk, Yusef-Zadeh:2004fk}, as Figure 1 illustrates.  This has led some to the notion that there is a pervasive vertical field throughout the central $\sim$100 pc that is ÒilluminatedÓ locally where relativistic electrons happen to be injected.  Those electrons are constrained to diffuse primarily along the field lines, thereby producing the NTFs.  The alternative is that the filaments are a local phenomenon, in which tubes of enhanced magnetic flux are produced by a phenomenon such as turbulence.  In the model of \citep{Boldyrev:2006fk}, for example, the enhanced-field filaments are confined by the effective pressure of large-scale turbulence.  In their model, the turbulent cells expulse the field, and concentrate it in regions between the cells.  A strong constraint of this model is that the helicity of the turbulence must be organized in such a way that the resulting NTFs are predominantly vertical.  Another local model for NTF production is that of \cite{Shore:1999vn}, in which the NTFs are interpreted as the plasma tails of molecular clouds in a Galactic wind.  In the context of this scenario, it is not evident how local production of NTFs can lead to orientations of most of the NTF population on scales of a few hundred parsecs.  

The other important clue to the nature of the NTFs is that all of them that have been adequately investigated appear to be undergoing an interaction with a molecular cloud somewhere along their length.  \cite{Serabyn:1994cr} used this fact, and the orthogonality of the magnetic field in NTFs and that inside of clouds, to argue that the relativistic electrons responsible for the radio emission are produced by magnetic field line reconnection at cloud surfaces.  The availability of free electrons is also an important element; the interaction sites of the clouds and the NTFs occur where the cloud surfaces have been ionized by a local stellar source of UV radiation.  Not only do the surface HII regions supply the electrons, but the turbulent ionization fronts also mix the external and internal magnetic field systems, thereby enhancing the rate of reconnection.  
Examples of such interfaces include the G359.1-0.2 and G359.54+0.18 NTFs \citep[respectively]{Uchida:1996zr, Staguhn:1998vn}, but there are numerous others.

A fraction of the NTFs does not conform to the uniform poloidal field picture.  A few relatively long filaments (10's of pc) are tilted by relatively large angles relative to the vertical to the Galactic plane, and a population of short NTFs appears to have more or less random orientations \citep{Lang:1999ly, LaRosa:2004kx, Yusef-Zadeh:2004fk}.  The randomly oriented filaments may represent a separate population produced by local events such as jets or supernova shocks that can deform the magnetic field.  Some of the non-conforming NTFs are clearly formed in this way.  A few strongly curved NTFs in the Sgr A radio complex, for example, have their centers of curvature facing Sgr A*, suggesting that collimated outflows from the central black hole or the region around it have deformed what might otherwise have been a uniformly vertical magnetic field \citep{Morris:2014hc}.  Figure 3 shows the most highly distorted example from that work.

\begin{figure}
    \centering
        \includegraphics[width=9cm]{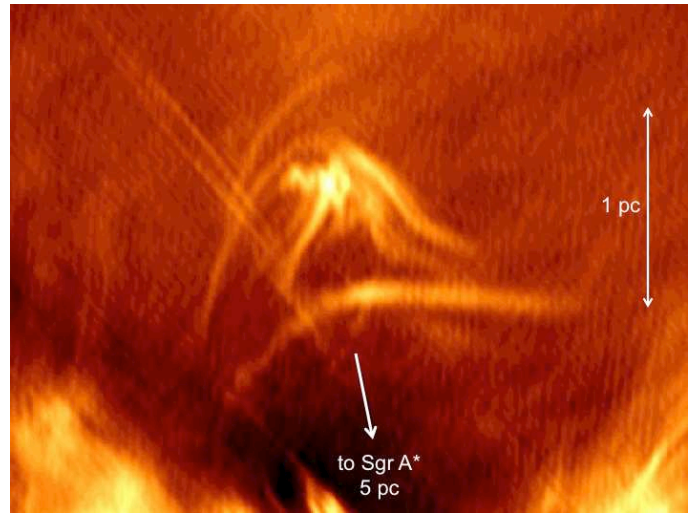}

        \vspace{-1mm}
        \caption{The "Northern Curl" radio source, as observed with the JVLA at 6cm \citep[and Zhao, Morris \& Goss 2014, in preparation]{Morris:2014hc}.  All of the emission in this image is nonthermal.  The direction to Sgr A* is indicated.  A similar, but larger feature is observed on the opposite side of Sgr A*.  The doubled filament that is superimposed on the strongly deformed magnetic structure is approximately parallel to the Galactic plane, and represents one of the small NTFs that does not conform to the presumed large-scale poloidal field.  } 
\end{figure}

The rigidity of the magnetic field in the NTFs, as suggested by their smoothly varying and large-scale curvatures and the absence of significant deformations, led to an estimate of milligauss field strength within them \citep{YZM1987a, YZM1987b}.  The argument rests on the fact that certain NTFs are clearly interacting with molecular clouds in which the internal velocity dispersion, $\Delta V$, is large -- a few tens of km/sec.  Yet despite the interaction, the NTFs are perturbed in only the most subtle ways, such as a bending by a few degrees or a slight, large-scale indentation at the site of an apparent cloud impact \citep[e.g.,][]{Tsuboi:1997fk}.  The magnetic field pressure thus must be at least comparable to the internal turbulent pressure, $\rho \Delta V^2$, of the clouds, where $\rho$ is the cloud mass density, {\em unless} the Alfv\'en speed is large enough to radiate away the perturbations to the field lines on sufficiently short time scales that they remain unobservable.  One potential concern about the milligauss field strength estimate is the implied short synchrotron lifetime of the electrons \citep[see][]{Ferriere:2009fk}, but \cite{Morris:2006fk} offers reasons why this need not be a fatal concern.  Another potential concern is that, if the milligauss field is pervasive throughout a region of radius R$_{gc}\sim$100 pc, the total magnetic energy is quite large, amounting to 4$\times$10$^{54}~B_{mg}^2$ (R$_{gc}$/100 pc)$^3$ ergs \citep{Morris:1994uq}.  However, this is not too much greater than the total energy content of the diffuse $\sim$10$^8$ K gas that might occupy the central few hundred parsecs of the Galaxy \cite[and references therein]{Koyama:2009nx, Koyama:2011ij}, and it is substantially less than the magnetic energy content of the overlying Fermi Bubbles \citep[see below][]{Carretti:2013uq}.  Whether the X-ray evidence can be ascribed to such a hot gas reservoir or to an unresolved collection of point sources \citep{Muno:2004fk, Revnivtsev:2009fv} remains one of the key questions in current Galactic center research.

On scales of a several hundred parsecs -- substantially larger than the region occupied by the NTFs -- there are constraints on the magnetic field strength provided by the diffuse radio synchrotron emission from cosmic rays, coupled with gamma ray emission up to TeV energies \citep{Crocker:2010kx}. \cite{Crocker:2011vn} find that the average field strength on such large scales lies within the range from 0.06 to 0.4 mG.  This is not grossly inconsistent with the estimates from NTFs, given the likelihood that the field diverges, and its strength thereby declines with distance, but it does raise the question of whether the milligauss fields have been overestimated, or whether the milligauss estimates apply only locally to the NTFs.  

\section{Coexistence of the Two Field Systems}
Soon after the NTFs were discovered \citep{Yusef-Zadeh:1984fk}, \cite{Uchida:1985tg} formulated a model in which gas in the differentially rotating Galactic disk interacts with a globally poloidal field by dragging it into a twisted, toroidal configuration. Inflow of gas toward the center along the field lines then leads to a contraction of the field, which, in turn, could be invoked to drive a helical galactic outflow toward the poles.  This model has inspired considerable discussion, and in particular has been invoked to explain how the field within intermediate-density clouds of the CMZ can be largely toroidal while the field outside of clouds above the Galactic plane appears to be poloidal \citep[e.g.,][]{Chuss:2003fk}.  However, the shapes of the NTFs do not support this hypothesis.  There is no evidence that the field lines are coupled strongly to the gas in the Galactic disk, or that the inertia of that gas has twisted the field lines away from a predominantly vertical orientation.   The Radio Arc, for example, punches through the Galactic plane without significant distortions being introduced (Figures 1 and 2).  Another example, the isolated radio "thread", G0.08+0.15 (shown in Figure 2), also passes through the gas layer without acquiring a component along the Galactic plane.  All evidence points to the conclusion that the field systems inside and outside of clouds are quite independent.  Clouds orbit through a poloidal field, perhaps suffering some magnetic reconnection activity at their surfaces which can, under the right circumstances lead to the illumination of the field lines \citep{Serabyn:1994cr}, and perhaps encountering some magnetic drag \citep{Morris:1994uq}, but not interacting strongly enough to affect the field geometry or the cloud dynamics, except on very long time scales.  Any momentum transferred from CMZ clouds to the mass-loaded field is carried away from the plane by Alfv\'en waves having wavelengths much larger than the scale of the CMZ.  It follows that clouds "slip through" the vertical field lines; there is less energy required to displace the intercloud field lines around an orbiting cloud than in compressing the intercloud field along the direction of the cloud's motion.  

For the field within CMZ clouds, we are left with the following scenario.  Because of dynamical friction \citep{Stark:1991dz}, plus magnetic drag, the residence time of CMZ material is limited to somewhat less than 10$^9$ years.  The gas is brought in to the CMZ from the Galactic disk by the action of the bar potential, and migrates inexorably inwards until it most likely forms stars (only a tiny fraction apparently reaches the central parsec), or is ejected in a Galactic wind.  Consequently, the magnetic field within CMZ clouds has been brought in with the gas, and is amplified by the differential rotation that quasi-continuously stretches it into tidal streams.  Eventually, the predominantly toroidal field within clouds equilibrates as some magnetic flux is carried away by the wind and some is lost to reconnection.  

The evidence for a Galactic wind is growing.  If diffuse hot gas is truly present at the Galactic center at a temperature near 10$^8$ K, then it is unbound and should lead to a bipolar Galactic wind \citep{Muno:2004fk, Belmont:2005fu}.  Indeed, 
\cite{Nishiyama:2010uq} invoke such a wind to account for the evolution of their inferred magnetic field vectors with Galactic latitude.
Even in the absence of an unbound, hot plasma, however, the fairly active star formation in the CMZ, accompanied by supernovae, can drive a Galactic fountain that carries magnetic flux out of the CMZ \citep{Crocker:2012kl}.  According to Crocker, cosmic rays associated with star formation and supernovae in the CMZ are advected out of the CMZ by the wind at the rate needed to account for the gamma-ray flux from the very-large-scale Fermi Bubbles \citep{Su:2010fk}.  A recent discovery by \cite{Carretti:2013uq} of polarized radio lobes apparently wrapping around the Fermi Bubbles has led them to interpret this in terms of magnetic flux that has been carried out from the CMZ by the wind.  The helical "wrapping" is ascribed to Galactic rotation, carrying the most recent source of the wind -- a site of major star formation -- around the Galactic center.  

The poloidal field could, in principle, be ascribed to shear in a Galactic wind rising out of the CMZ, but such a model would be unable to account for its strength.  The vertical field at the Galactic center can instead be reasonably well modeled in terms of the concentration of the Galaxy's primordial field, driven by the inexorable inflow of gas toward the center over the lifetime of the Galaxy \citep{Chandran:2000fk}.  In that case, the poloidal and toroidal fields at the Galactic center are completely independent of each other.  Many questions, remain, however, such as where and how the poloidal field transitions into the dominant toroidal field at some point in the disk, and how that might relate to the X1-X2 orbital transition near the inner Lindblad resonance \citep{Morris:1996oq}.

\bibliography{GCB}{}

\end{document}